\documentclass[pre,normal]{revtex4}

\newcommand{\lgl}{\langle}
\newcommand{\rgl}{\rangle}

\begin{document}

\title{On a suggested link between irreversibility and causality}
\author{W. \surname{Pietsch}}
\address{Department for Philosophy of Science, University of Augsburg, Universit\"atsstrasse 10, D-86135 Augsburg, Germany}

\begin{abstract}
It has been a long time issue in statistical physics how to combine reversible microscopic equations with irreversible macroscopic behavior. Recently, Evans and Searles have suggested causality as the key concept for a solution to the irreversibility problem [D.\ J.\ Evans and D.\ J.\ Searles,\ Adv.\ Phys.\ {\bf 51},\ 1529 (2002) and Phys.\ Rev.\ E {\bf 53},\ 5808 (1996)]. This proposal is examined from a philosophical, i.e.\ conceptual, perspective, which reveals that the point of view taken by Evans and Searles is identical with the one already suggested by Boltzmann. According to Boltzmann initial conditions are responsible for the observed irreversibility of the macroscopic world. It is shown that reversibility requires a concept of causality, where causes and effects are interchangeable -- at least on probabilistic grounds. Causality as interpreted by Evans and Searles is not compatible with microscopic reversibility and should itself be derived in order to solve the irreversibility problem.
\end{abstract}

\maketitle

\pagestyle{headings}

\section{Introduction}
Recent times have shown considerable advance in the field of nonequilibrium thermodynamics, also called thermodynamics of irreversible processes. This important progress is closely connected with two related theorems: the Fluctuation Theorem \cite{evans93,evans96,evans02} and the Jarzynski Relation \cite{jarzynski96,jarzynski97,jarzynski00}. In 1999 Crooks showed that the latter is in fact a special case of the former \cite{crooks99}. By making exact statements about irreversible processes these relations touch fundamental issues in physics in general -- for example they address the still partly unresolved question about the origin of irreversibility in the macroscopic world. In the next section, I will outline this problem, which from now on I shortly call irreversibility problem. The impact on several philosophical questions in physics, i.e.\ those dealing with the conceptual basics, was clearly realized by Evans and Searles \cite{evans96,evans02}. In their work related to the Fluctuation Theorem, they suggest the concept of causality as a solution to the irreversibility problem. In the following text, I will analyze the use of causality in \cite{evans96, evans02} from a philosophical perspective. Subsequently, I will briefly sketch my own point of view concerning the connection between causality and irreversibility.

\section{Deriving Irreversible Thermodynamics from Reversible Microscopic Equations}

\subsection{An outline of the irreversibility problem and the standard solution by Boltzmann}
\label{sec}
It is only in respondence to the criticism brought forward by Poincar\'e, Loschmidt, Zermelo and others, that Boltzmann arrived at a clear standing concerning a microscopic derivation of the Second Law of Thermodynamics. Poincar\'e famously stated, that he cannot believe in a derivation when the results contradict the assumptions. Loschmidt formulated the reversibility objection, which reads as follows (not in the most general, but perhaps in the clearest form): When at a certain point in time, we reverse the velocities of all particles, then the system will develop as though backwards in time. This means in particular, that if entropy is increasing before the reversal then after the reversal entropy will decrease and vice versa. Since the \emph{reversible} microscopic equations allow both processes, it seems that nature has no preference for entropy--increasing or entropy--decreasing processes. Boltzmann's solution to this problem is that entropy--increasing processes are much more probable than entropy--decreasing processes -- thus identifying the increase of entropy with the development of a system towards more probable states. These statements about the probability of physical processes cannot come from the microscopic equations, since those are deterministic and they do not make any statements about probabilities. 

Thus, according to Boltzmann probability is the key concept responsible for introducing irreversibility into physics. This essentially refers to the probability of initial conditions. Obviously, this classic solution is closely related to the interpretation of entropy as probability, and thus has to be considered still valid today. However, the problem of irreversibility is not fully resolved at this point. Many questions remain. The most obvious was posed (and immediately dismissed) by Boltzmann himself: `Why do we live in such an improbable world?' In a reply to Zermelo, he writes that we cannot expect to find an answer to this question just as we cannot answer, why there are phenomena and laws of nature at all \cite{boltzmann96}. 

Many philosophers of science and physicists are not fully satisfied with this remark. Also, there are other important questions remaining. Some of them have come up only in the time after Boltzmann -- e.g. if the cosmological arrow of time, which is determined by the expansion of the universe, plays a role for irreversibility. Also, it has often been questioned, whether microscopic physics is really fully reversible. A famous example in this respect is the debate between Einstein and Ritz concerning the reversibility of electrodynamics, particularly of radiation phenomena \cite{einstein09}. Another discussion concerns the irreversibility of quantum mechanics. To some the measurement--process seems to be a good candidate for implementing irreversibility into the fundament of physics. Other interesting approaches to the irreversibility problem have stressed the information-theoretical aspects of entropy \cite{jaynes57}, the openness of systems and the influence of boundary conditions \cite{hoover98}, or the dynamic instability in chaos theory \cite{prigogine79}. (For a good overview over the whole topic see \cite{zeh}.)

\subsection{The ansatz by Evans and Searles}
In the following I examine the solution to the irreversibility problem as suggested by Evans and Searles in \cite{evans96} and repeated in \cite{evans02}. Evans and Searles start with the response of a system to an external force,
\begin{equation}
\delta B(t_1)=L(t_1-t_2)F(t_2)\delta t_2.
\end{equation}
$F(t_2)$ is the external force applied to the system at time $t_2$. $\delta B(t_1)$ is the linear response at time $t_1$ of a system characterized by a response function $L(t_1-t_2)$. This system is invariant to time translation. The total response at time $t$ due to an external force at all previous times is
\begin{equation}
B_C(t)=\int_{-\infty}^{t} L_C(t-t_1)F(t_1)dt_1.
\label{eq:bc}
\end{equation}
As a counterpart to Eq.\ (\ref{eq:bc}), which they call a causal response, Evans and Searles define an anticausal response
\begin{equation}
B_A(t)=-\int_t^{+\infty} L_A(t-t_1)F(t_1)dt_1.
\label{eq:ac}
\end{equation}
Their argument can be summed up as follows: We live in a time-reversible world, hence all processes happen in both time-directions. So, there must be cases, where effect precedes cause. These cases are `never' observed in the macroscopic world. However, on the microscopic level they have a certain probability. From these microscopic cases, Evans and Searles derive their concept of an anticausal world.
 
This is problematic, since all processes in our reality, which is equipped with a determined direction of time, are causal by definition. Causality in this case is defined by the fact, that effects are \emph{preceded} by causes. So, those supposedly `anticausal' microscopic processes are by definition also causal, since they belong to the reality, we experience. Essentially, Evans and Searles implicitly \emph{define} their anticausal world as a world where just the probabilities of all processes are reversed: 
\begin{equation}
P_C(A,t_1;B,t_2)\equiv P_A(B,t_3;A,t_4),  
\label{eq:defanti}
\end{equation}
with fixed times $t_1$ to $t_4$ and $t_2-t_1=t_4-t_3$. $P_C(A,t_1;B,t_2)$ is the probability, that a system is in state A at time $t_1$ and in state B at time $t_2$. All probabilities $P_A$ relate to the postulated anticausal world, all probabilities $P_C$ to reality. Ultimately, the anticausal world as defined by Evans and Searles would correspond to the universe seen under time-reversal. Since the real universe is governed by reversible laws, the anticausal universe is also governed by reversible laws. The difference between the causal and the anticausal universe is simply that initial conditions of processes in the real universe become final conditions of corresponding processes in the time-reversed universe and vice-versa. 

Through the definition of their anticausal world, Evans and Searles link the irreversibility of the world we experience to the difference between initial and final conditions. Again, I employ the word `define', because the anticausal processes do not exist in reality. In the end, Evans and Searles arrive at the same conclusion as Boltzmann. As mentioned above, Boltzmann stated that there is no answer to why we live in such an improbable world. Evans and Searles also do not address this problem.      

Consider now a finite causal process and its corresponding anticausal equivalent \cite{evans96}:
\begin{equation}
B_C(t)=\int_{0}^{+t} L_C(t-t_1)F_C(t_1)dt_1,
\label{eq:bc1}
\end{equation}
\begin{equation}
B_A(-t)=-\int_{-t}^{0} L_A(-t-t_1)F_A(t_1)dt_1.
\label{eq:ac1}
\end{equation}
Eq.\ (\ref{eq:ac1}) describes just the time-reversed process (\ref{eq:bc1}). Following from Eq.\ (\ref{eq:defanti}) with $t_1=0$, $t_2=+t$, $t_3=-t$, and $t_4=0$, we have for the time-reversed process, resulting from the exchange of initial and final states:
\begin{equation}
F_C(s)=\pm F_A(-s) \equiv v_1 F_A(-s), \quad \forall\, s \textnormal{\ with } 0\leq s \leq t.
\end{equation}
The sign depends, if the generalized force $F_C$ is symmetric or antisymmetric under time reversal. We also have
\begin{equation}
B_C(t)=\pm B_A(-t)\equiv v_2 B_A(-t).
\label{eq:bca}
\end{equation}
Again, the sign depends, if the quantity $B_C$ describing the response of the system is symmetric or antisymmetric in time. From Eqs.\ (\ref{eq:bc1}) and (\ref{eq:ac1}) in combination with (\ref{eq:bca}) we can derive
\begin{equation}
\int_{0}^{+t} L_C(t-t_1)F_C(t_1)dt_1=-v_2\int_{-t}^{0} L_A(-t-t_1)F_A(t_1)dt_1=-v_2\int_{0}^{+t} L_A(-t+t_1)F_A(-t_1)dt_1,
\end{equation}
which gives
\begin{equation}
\int_{0}^{+t} (L_C(t-t_1)+v_2 v_1 L_A(-t+t_1))F_C(t_1)dt_1=0.
\end{equation}
Since this must hold for all processes, we have
\begin{equation}
L_C(s)=-v_2 v_1 L_A(-s).
\label{eq:acbc}
\end{equation}
In their derivation in \cite{evans96}, Evans and Searles explicitly consider Green-Kubo relations for shear viscosity
\begin{equation}
\frac{dJ_{\bot}(k_y,t)}{dt}=\frac{-k_y^2}{\rho}\int_0^t\eta_C(k_y,t-s)J_{\bot}(k_y,s)ds, \quad t>0
\label{eq:relc}
\end{equation}
and the anticausal equivalent 
\begin{equation}
\frac{dJ_{\bot}(k_y,t)}{dt}=\frac{k_y^2}{\rho}\int_t^0\eta_A(k_y,t-s)J_{\bot}(k_y,s)ds, \quad t<0.
\label{eq:rela}
\end{equation}
$\eta_C$ and $\eta_A$ are the response functions (also called memory functions). $J(k_y,t)$ is the wave-vector dependent transverse momentum density. For interpretational details, that are not of importance in this context, compare \cite{evans96,evans90}. It is clear, that for the specific example, we have 
\begin{equation}
v_1=-v_2,
\end{equation}
because $\frac{dJ_{\bot}(k_y,t)}{dt}$ is the time-derivative of $J_{\bot}(k_y,s)$. From this follows directly using Eq.\ ($\ref{eq:acbc}$)
\begin{equation}
\eta_C(k_y,t)=\eta_A(k_y,-t),
\label{eq:result}
\end{equation}
which is the result derived in Section III of \cite{evans96}. 

In their proof, Evans and Searles implicitly make the assumption (\ref{eq:defanti}), when they identify anticausal autocorrelation functions with causal autocorrelation functions and thereby determine the probabilities of anticausal processes. Eq.\ (\ref{eq:result}) can be written out as \cite{evans96}: 
\begin{equation}
\frac{V}{k_BT} \lgl P_{yx}(t) P_{yx}(0) \rgl_C\equiv \frac{V}{k_BT} \lgl P_{yx}(-t) P_{yx}(0) \rgl_A, \quad t\geq 0.
\label{eq:help}
\end{equation}
Note that contrary to Evans and Searles we distinguish between causal and anticausal autocorrelation functions. For the exact definition of $P_{yx}$, that is again not important in this context, compare \cite{evans96,evans90}. Eq.\ (\ref{eq:help}) translates to
\begin{equation}
\int \int dP'_{yx} dP_{yx} P'_{yx} P_{yx} f_C(P'_{yx},t; P_{yx},0)dt\equiv \int \int dP'_{yx} dP_{yx} P'_{yx} P_{yx} f_A(P'_{yx},-t; P_{yx},0)dt.
\end{equation}
To connect all processes and to fully define the anticausal world, this must hold for all autocorrelation functions, i.e.
\begin{equation}
f_C(P'_{yx},t; P_{yx},0)\equiv f_A(P'_{yx},-t; P_{yx},0).
\label{eq:fac}
\end{equation}
$f_C(P'_{yx},t; P_{yx},0)$ is the probability distribution for the system being in state $P'_{yx}$ at time $t$ and state $P_{yx}$ at time $0$. $f_A(P'_{yx},-t; P_{yx},0)$ is the probability distribution for anticausal processes. Eq.\ (\ref{eq:fac}) corresponds to Eq.\ (\ref{eq:defanti}). It is essentially a relation between the initial states of an ensemble of causal processes and the final states of the corresponding ensemble of anticausal processes and vice versa. Evans and Searles implicitly make the assumption (\ref{eq:defanti}) by not distinguishing between causal and anticausal autocorrelation functions. Thereby they determine the probability distributions for anticausal processes.

To sum up, the argument in \cite{evans96, evans02} is that the origin of the observed irreversibility in the world has to do with causality. This program is carried out by comparing the real world with a proposed anticausal world. The main property of this anticausal world is the implicit assumption, that compared to the real causal world initial and final conditions are exchanged. Evans and Searles, however, do not address the question, why the initial conditions in the real world are such as observed. Thereby, Evans and Searles arrive simply at the standard conclusion concerning the irreversibility problem. This solution was already proposed by Boltzmann (cp.\ for example \cite{boltzmann96}). In \cite{evans96} Evans and Searles just rename this interpretation, that irreversibility stems from the difference between initial and final conditions. Their expression is causality. 

\subsection{The role of the Fluctuation Theorem}
The Fluctuation Theorem does not change the conclusion we have just drawn from \cite{evans96}. The Fluctuation Theorem according to Evans and Searles `gives an analytical expression for the probability of observing Second Law violating dynamical fluctuations in thermostatted dissipative non-equilibrium systems' \cite{evans02}. It shall only be noted that these processes do not violate the Second Law in the probabilistic sense, in which it must be understood in the consequence of the work done by Boltzmann. But this is only a matter of names. More importantly the Fluctuation Theorem is fully compatible with Boltzmann's view that systems develop with greater probability into states closer to equilibrium (i.e.\ the most probable state) than in the opposite direction. Note, that this probabilistic interpretation of the Second Law as advocated by Boltzmann, Planck, Einstein and others makes no assumptions regarding if the systems are of finite or infinite size and if processes have finite or infinite duration. The systems are only assumed to be isolated.  

So, time-asymmetry always originates in the fact, that we start a process far from equilibrium, i.e.\ that we start from an improbable state. When `deriving' the Second Law, it must always be assumed that the initial state is less probable than the final state. This is a necessary condition as long as one clings both to a probabilistic interpretation of entropy $S=k_B\ln \Omega$ and reversible microscopic equations. Evans and Searles accept the two of them.  

\section{The Concept of Causality Required by Reversibility}
The concept of causality employed by \cite{evans96,evans02} in the suggested solution to the irreversibility problem is one that is often used in physics and also philosophy, but which is not suitable for this specific context. It uses a definite time-order, where the time-coordinate of the cause is always smaller than (or maybe equal to) the time-coordinate of the effect and where cause and effect are not generally exchangeable. 
This concept of causality is used in linear response theory. There, an outside force determines through a response function the reaction of the system. The outside force is considered the cause, the reaction is the effect. Causality in this context means that the reaction of the system at time $t$ may only be influenced by outside forces at times $t'\leq t$. 

We will now examine causality in Newtonian physics and find that there a totally different concept is employed, which does not allow for a distinction between cause and effect.
The third Newtonian axiom `actio=reactio' states the equivalence of the acting and the reacting force. In particular this means that one cannot distinguish in principle, which is the acting and which the reacting force, i.e.\ cause and effect cannot be distinguished in principle. Consider two massive particles A and B acting upon each other. The properties of particle A, e.g.\ its mass, are responsible for the motion of particle B and vice versa. It cannot be determined `which acceleration of which particle comes first'. In case of Newtonian physics there is no distinction between cause and effect through time-order, as cause and effect are happening simultaneously (velocity of interaction is infinite). This can be seen very clearly in Newton's law of gravity
\begin{equation}
F=G\frac{m_A m_B}{r^2}.
\end{equation}
The physical origin for the fact, that cause and effect are interchangeable, is the identity of passive and active mass for every particle. This corresponds to the third Newtonian axiom. Thus, in Newton's theory there is only \emph{inter-}action. Cause and effect always have the same time-coordinate and are exchangeable.

Accepting the conceptual equality of cause and effect directly leads to the reversibility of all microscopic equations and vice versa.
In fact, the Newtonian laws are only time-reversible since they rely on a concept of interaction and not time-ordered causation. This holds for the Hamiltonian formulation of mechanics as well as the equivalent Newtonian formulation. 

In Maxwellian electrodynamics, the situation might at first seem different, since in the framework of this theory signals can travel at most with the speed of light. Thus, at first sight there seems to be a fixed time-order of causes and effects. Still, Maxwell Equations are known to be invariant under time reversal and simultaneous reversal of the magnetic field. It is the existence of both advanced and retarded solutions that makes Maxwellian electrodynamics invariant under time reversal. When advanced solutions are dismissed on the basis of `causality', as is done in many books on electrodynamics, already the reversibility of the theory is given up. So, the reversibility of Maxwellian electrodynamics depends on the acceptance of advanced solutions, in other words solutions where in a conventional dictation cause and effect would be exchanged. Only under this premise is the theory time-reversible because only then it is not possible to distinguish between cause and effect because every process can happen in both directions: For every process with cause A and effect B one must accept the possibility of a process with cause B and effect A. In addition we do not make any assumptions about the probabilities of those processes (which in fact is not possible in the framework of Maxwellian electrodynamics). 

If we speak of time-ordered causality, we do not consider reversible causality as in electrodynamics but imply that cause and effect are not conceptually equivalent. In a strong version of time-ordered causality at least for some processes with cause A and effect B there does not exist a corresponding process with cause B and effect A. In a weaker version of time-ordered causality, for every process with cause A and effect B there exists a corresponding process with cause B and effect A, but the probabilities of both processes differ. The weaker version is essentially the solution to the irreversibility problem as given by Boltzmann. This solution is possible because the microscopic equations do not make any statements about the probabilities of processes. They only state, if processes are allowed in principle or not.

It is tautological to say that time-ordered causality can solve the problem of irreversibility. The microscopic equations do not exhibit time-ordered causality. Once we admit a clear distinction between events of cause and events of effect, then of course physics is not reversible by definition. In \cite{evans02} the strong version of time-ordered causality is suggested by the cited example of throwing on a switch and `afterwards' observing the light go on. Then, like in Boltzmann's work the weak version is employed to solve the irreversibility problem. 

As for the argument in \cite{evans96,evans02}, we have found that reversibility requires a concept of causality, where a distinction between causes and effects cannot be established -- neither on deterministic nor on probabilistic grounds (corresponding to the strong or weak version of time-ordered causality). Full reversibility and a concept of causality, where causes and effects are interchangeable, are two sides of the same coin. Both statements are equivalent. When Evans and Searles give the example of switching on a light, they presumed a concept of causality, that is not compatible with reversibility. However, Evans and Searles have not shown how their concept of causality derives from the concept of causality necessary for reversible microscopic equations - other than by making assumptions about initial conditions just as Boltzmann did more than a hundred years ago. All questions that go beyond Boltzmann's solution -- e.g.\ the ones of Section \ref{sec} -- remain unanswered.

\end{document}